\documentclass{aa}  
\usepackage{graphicx}
\usepackage{txfonts}
\def\a{{$\alpha$}}

\newcommand{\h}{$^{\rm h}$}
\newcommand{\m}{$^{\rm m}$}
\newcommand{\s}{$^{\rm s}$}
\newcommand{\dd}{$\delta$}
\newcommand{\ha}{\rm H$\alpha$}
\newcommand{\hbeta}{\rm H$\beta$}
\newcommand{\HII}{\ion{H}{ii}}
\newcommand{\hnii}{{\rm H}$\alpha+[$\ion{N}{ii}$]$}
\newcommand{\nii}{$[$\ion{N}{ii}$]$}
\newcommand{\sii}{$[$\ion{S}{ii}$]$}
\newcommand{\oi}{$[$\ion{O}{i}$]$}

\newcommand{\oiii}{$[$\ion{O}{iii}$]$}
\newcommand{\snr}{\rm supernova remnant}

\newcommand{\et}{et al.}
\newcommand{\fluxa}{$10^{-16}$ erg s$^{-1}$ cm$^{-2}$ arcsec$^{-2}$}
\newcommand{\fluxb}{$10^{-17}$ erg s$^{-1}$ cm$^{-2}$ arcsec$^{-2}$}
\newcommand{\dens}{\rm cm$^{-3}$}
\newcommand{\vel}{\rm km s$^{-1}$}
\newcommand{\sulfur}{[S~{\sc ii}]}
\newcommand{\nitrogen}{[N~{\sc ii}]}
\newcommand{\oxygen}{[O~{\sc iii}]}
\newcommand{\siirat}{$[$\ion{S}{ii}$]\lambda\lambda\ 6716/6731$} 
\newcommand{\HNII}{{\rm H}$\alpha+$[N {\sc ii}]~6548~\&~6583~\AA}  
\newcommand{\OIII}{$[$\ion{O}{iii}$]$~5007~\AA}
\newcommand{\SII}{$[$\ion{S}{ii}$]$~6716~\&~6731~\AA}
\begin{document}
\title{First optical detection from the supernova remnant G 15.1$-$1.6}


\author{P. Boumis\inst{1}
\and J. Alikakos\inst{1,2}
\and P. E. Christopoulou\inst{2}
\and  F. Mavromatakis\inst{3}
\and E. M. Xilouris\inst{1}
\and C. D. Goudis\inst{1,2}
}

\offprints{P. Boumis}

\authorrunning{P. Boumis et al.}
\titlerunning{First optical detection from the SNR G 15.1$-$1.6}

\institute{ Institute of Astronomy \& Astrophysics, National
Observatory of Athens, I. Metaxa \& V. Paulou, P. Penteli, GR-15236
Athens, Greece.\\
\email{[ptb;johnal;xilouris;cgoudis]@astro.noa.gr}
\and Astronomical Laboratory, Department of Physics, University of
Patras, 26500 Rio-Patras, Greece.\\
\email{pechris@upatras.physics.gr}
\and Technological Education Institute of Crete, General Department of
Applied Science, P.O. Box 1939, GR-710 04 Heraklion, Crete, Greece.\\
\email{fotis@physics.uoc.gr}          
}

\date{Received 28 November 2007 / Accepted 17 January 2008}

\abstract{Deep optical CCD images of the supernova remnant G
 15.1$-$1.6 were obtained and filamentary and diffuse emission has
 been discovered. The images, taken in the emission lines of \hnii,
 \sulfur~and \oiii, reveal filamentary and diffuse structures all
 around the remnant. The radio emission at 4850 MHz in the same area
 is found to be well correlated with the brightest optical filaments.
 The IRAS 60$\mu$m emission may also be correlated with the optical
 emission but to a 
 lesser extent. The flux calibrated images suggest that the optical
 emission originates from shock-heated gas (\sulfur/\ha\ $>$ 0.4),
 while there is a possible \HII\ region (\sulfur/\ha\ $\sim$ 0.3)
 contaminating the supernova remnant's emission to the
 east. Furthermore, deep long--slit spectra were taken at two bright
 filaments and also show that the emission originates from shock
 heated gas. An \oiii\ filamentary structure has also been detected
 further to the west but it lies outside the remnant's boundaries and
 possibly is not associated to it. The \oiii\ flux suggests shock
 velocities into the interstellar $``$clouds'' $\sim$100 \vel, while
 the \siirat\ ratio indicates electron densities up to $\sim$250
 cm$^{-3}$. Finally, the \ha\ emission has been measured to be
 between 2 to 7 $\times$ \fluxa, while the lower limit to the distance
 is estimated at 2.2 kpc.
}

\maketitle
%

\section{Introduction}

Supernova remants (SNRs) play an important role to understand the SN
mechanism, the interstellar medium (ISM) and their interaction. Most
of the SNRs have been detected in radio from their non--thermal
synchrotron emission. Observations of SNRs in X--rays allow us to
directly probe the hot gas inside the primary shock wave, while
optical observations offer an important tool for the study of the
interaction of the shock wave with dense material found in the
ISM. New searches in optical waveband continue to identify Galactic
SNRs (e.g. Boumis et al. \cite{bou02}, \cite{bou05}; Mavromatakis et
al. \cite{mav02}, \cite{mav05}) while in the last decade, observations
in X--rays have also detected new Galactic SNRs (e.g. Seward et
al. \cite{sew95}; see also Green \cite{gre06} for a complete catalogue).

G 15.1--1.6 is not a well known SNR, and was first detected by Reich
et al. (\cite{rei88}) in the Effelsberg 2.7 GHz survey and the radio
image was published by Reich et al. (\cite{rei90}). It is classified
as a shell--type SNR, with a spectral index of $\sim$0.8. Its angular
size is 30\arcmin $\times$ 24\arcmin and using the
brightness--to--diameter ($\Sigma$--D) relationship the distance of
the remnant was calculated at 5.7 kpc (Green \cite{gre06}). Radio
surveys of the surround area do not reveal any pulsar to by associated
with G15.1--1.6 while it has not been detected optically in the past.

In this paper we report the optical detection of G 15.1--1.6. We
present images of the remnant in the \hnii, [S~{\sc ii}] and [O~{\sc
iii}] emission lines. Deep long slit spectra were also acquired in a
number of selected areas. In Sect. 2 we present information about the
observations and data reduction, while the results of the imaging and
spectroscopic observations are given in Sect. 3. In Sect. 4 we discuss
the optical properties of this SNR, while in Sect. 5 we summarize the
results of this work.

\section{Observations}
A summary and log of our observations is given in
Table~\ref{table1}. In the subsections below, we describe the details of
these observations.

\subsection{Imaging}

\subsubsection{Wide--field imagery}

\begin{figure*}
\centering
\includegraphics[bb=50 20 575 600]{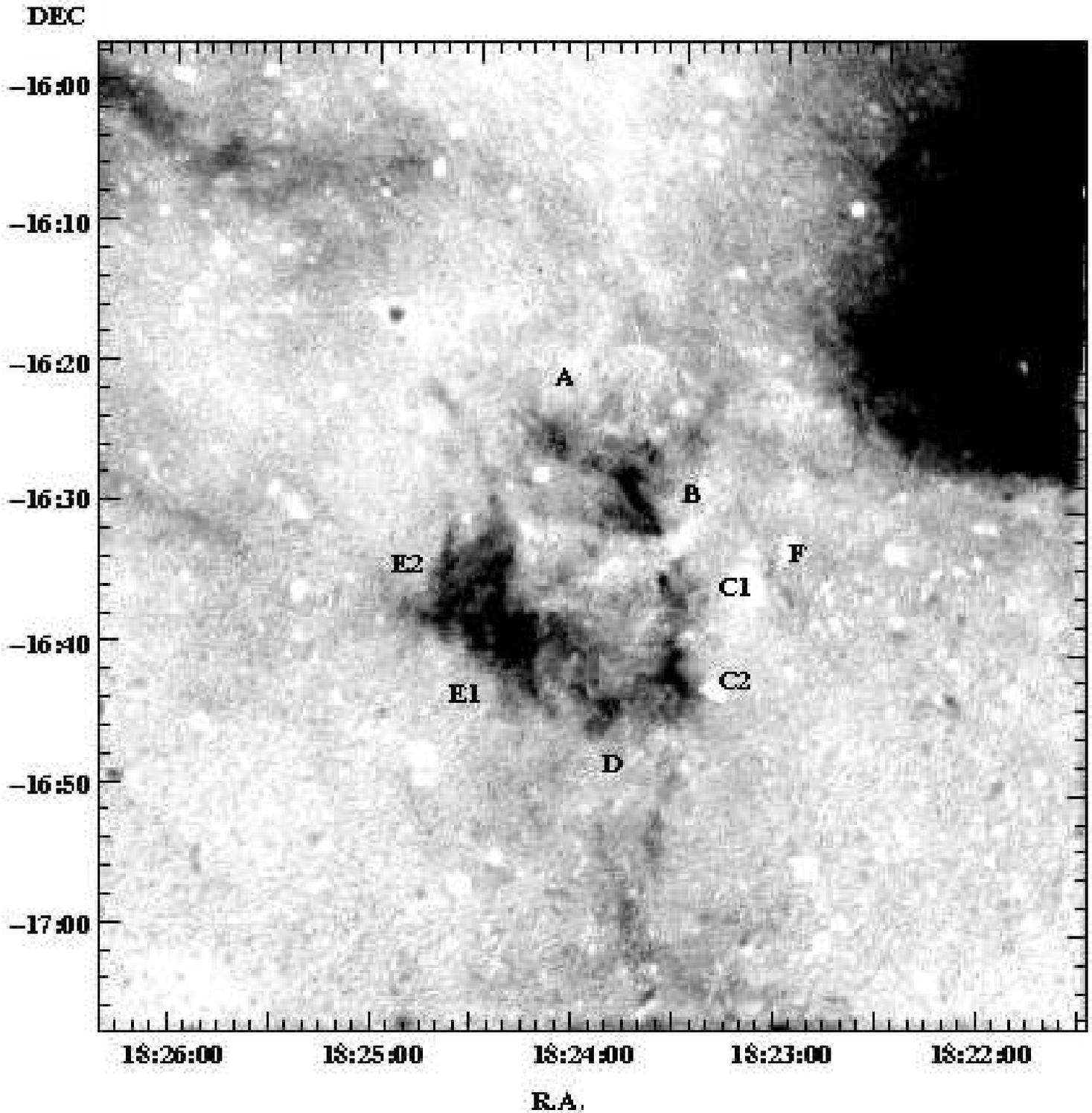}
\caption{The G 15.1$-$1.6 in the H$\alpha$ $+$ [N {\sc ii}]
filter. Shadings run linearly from 0 to 220$\times 10^{-17}$~erg
s$^{-1}$ cm$^{-2}$ arcsec$^{-2}$. The image has been smoothed to
suppress the residuals from the imperfect continuum subtraction. The
bright region in the north--west edge of the image is a known H {\sc
ii} region (Lockman \cite{loc89})}
\label{fig1}
\end{figure*}

\begin{figure*}
\centering
\includegraphics[bb=50 20 575 600]{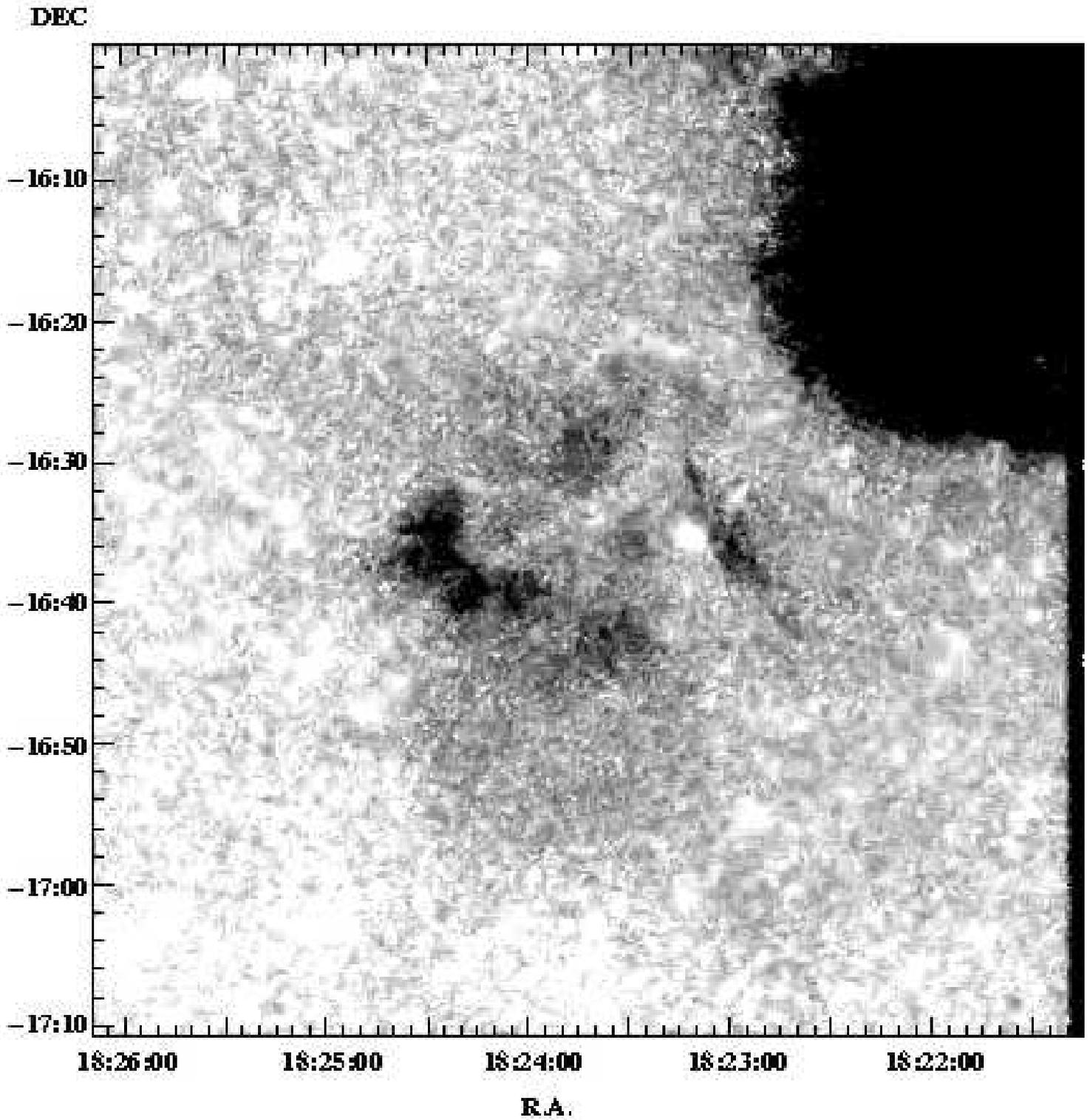}
\caption{The G 15.1$-$1.6 in the \oiii\
filter. Shadings run linearly from 0 to 45$\times 10^{-17}$~erg
s$^{-1}$ cm$^{-2}$ arcsec$^{-2}$. The image has been smoothed to
suppress the residuals from the imperfect continuum subtraction.}
\label{fig2}
\end{figure*}

G 15.1--1.6 was observed with the 0.3 m Schmidt--Cassegrain (f/3.2)
telescope at Skinakas Observatory, Crete, Greece in June 11, and
August 27, 28, and 30, 2005. The data were taken with a 1024 $\times$
1024 Thomson CCD with a pixel size of 19 $\mu$m resulting in a
70\arcmin\ $\times$ 70\arcmin\ field of view and an image scale of
4\arcsec\ per pixel. The area of the remnant was observed with the
\hnii, [S~{\sc ii}] and [O~{\sc iii}] filters. The exposure time was
set at 2400 sec for each observation and at continuum red and blue
filters the exposure time was 180 s. The continuum subtracted images
of the \hnii\ and [O~{\sc iii}] emission lines are shown in
Figs.~\ref{fig1} and \ref{fig2} respectively.

The IRAF and MIDAS packages were used for the data reduction. All
frames were bias subtracted and flat--field corrected using a series
of twilight flat--fields. Using the continuum images multiply with a
proper factor, we subtracted the stars in order to present the
remnant. For the absolute flux calibration, the spectrophotometric
standards stars HR5501, HR7596, HR7950, HR8634 and HR9087 (Hamuy et
al. \cite{ham92}) were used. The astrometric solution for all data
frames were calculate using the Hubble Space Telescope (HST) Guide
Star Catalogue (Lasker et al. \cite{las99}). All the equatorial
coordinates quoted in this work, refer to epoch 2000.

\subsubsection{High--resolution imagery}

\begin{figure*}
\centering
\includegraphics[bb=50 20 575 651]{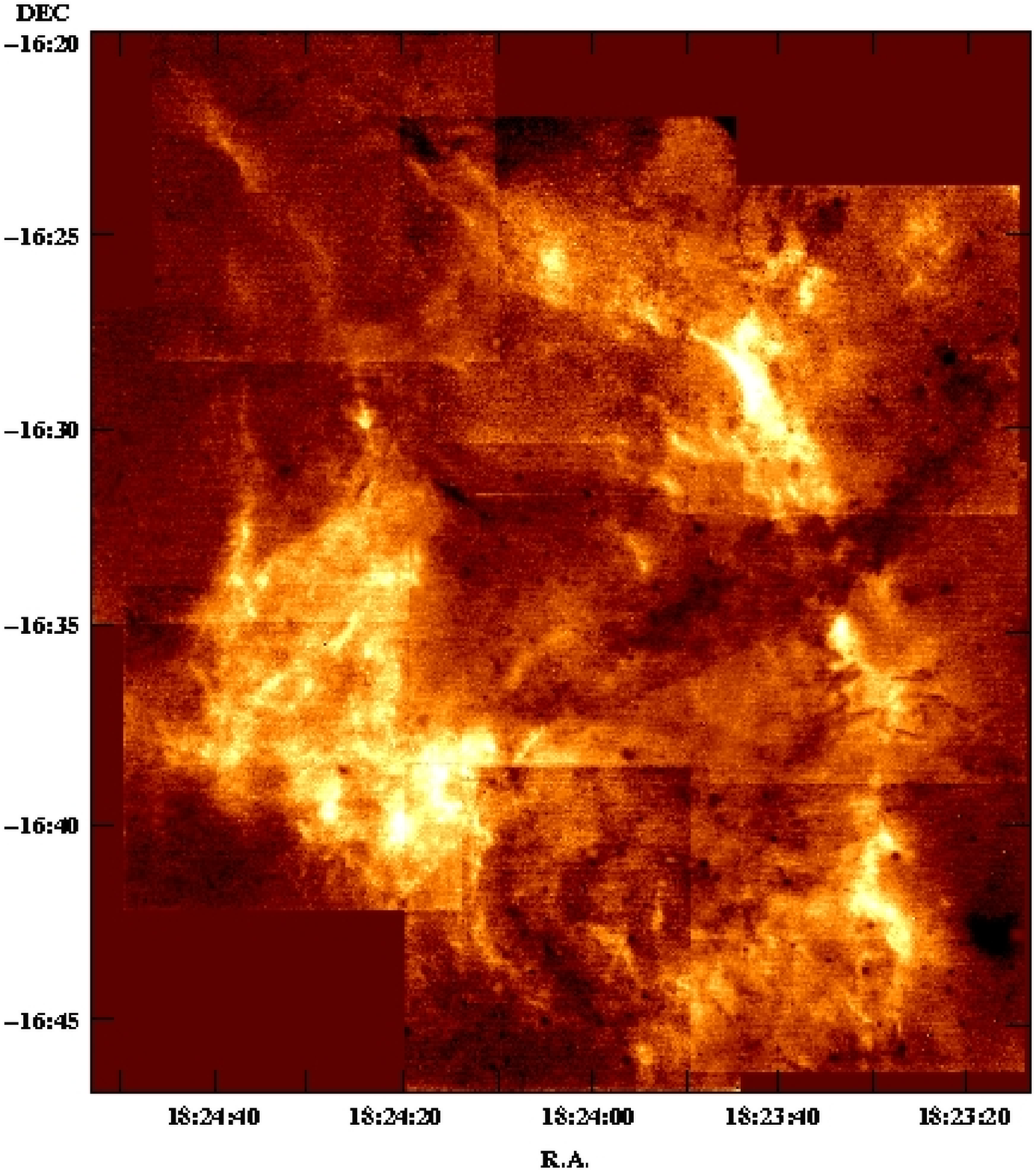}
\caption{The continuum--subtracted mosaic of G 15.1$-$1.6 taken with
the 1.3--m telescope in the light of H$\alpha$$+$[N {\sc ii}]. The
image has been smoothed to suppress the residuals from the imperfect
continuum subtraction.}
\label{fig3}
\end{figure*}

\begin{figure*}
\centering
\includegraphics[bb=50 20 575 665]{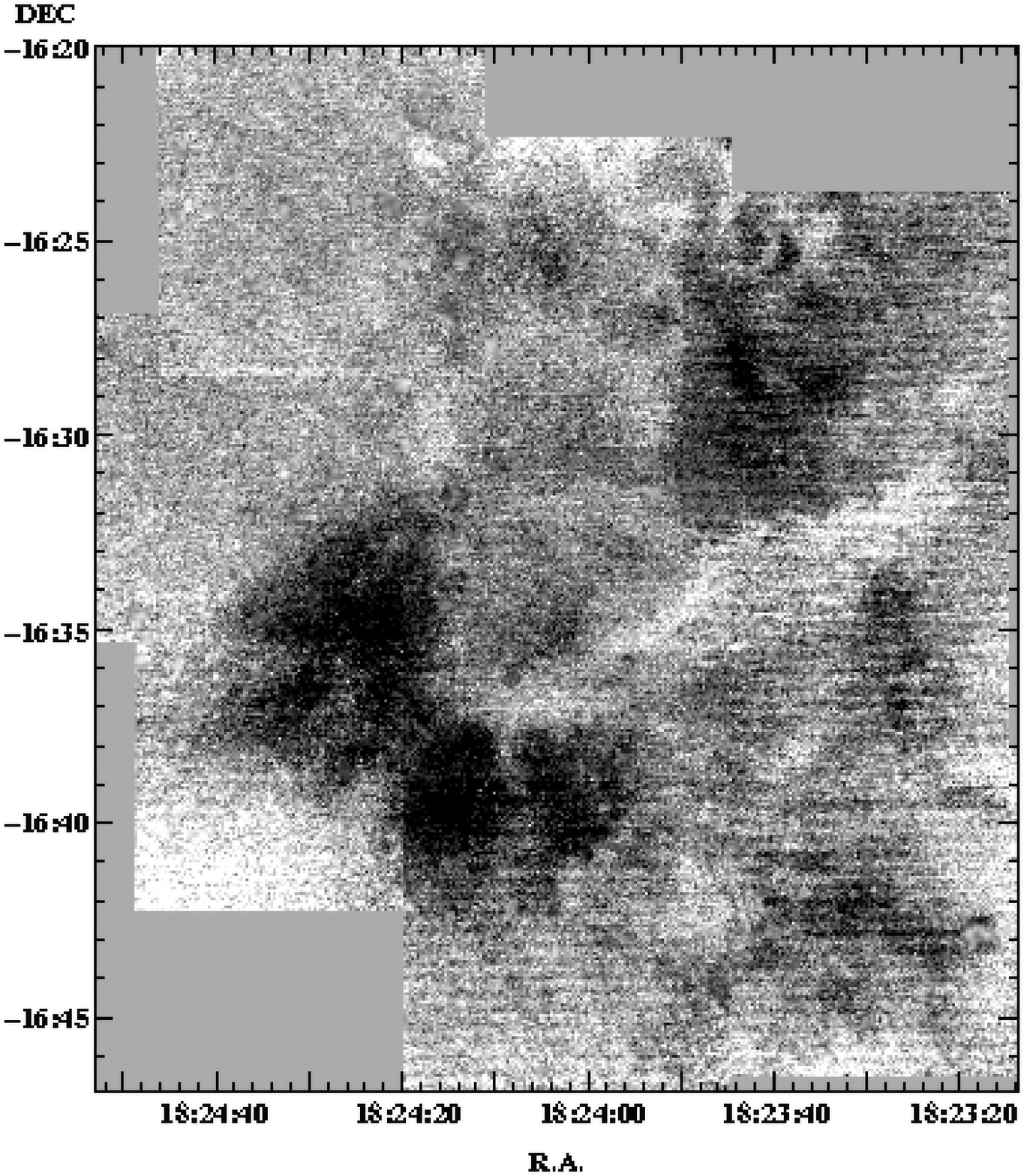}
\caption{The continuum--subtracted mosaic of G 15.1$-$1.6 taken with
the 1.3--m telescope in the light of \oiii. The image has been smoothed to
suppress the residuals from the imperfect continuum subtraction.}
\label{fig4}
\end{figure*}

Optical images at higher angular resolution of G 15.1$-$1.6 were also
obtained with the 1.3 m (f/7.7) Ritchey-- Cretien telescope at
Skinakas Observatory in July 4--7 and 8--10, 2007, using the
H$\alpha$+[N {\sc ii}] and the \oiii\ interference filters,
respectively. The detector was a 1024 $\times$ 1024 SITe CCD with a
field of view of 8.5 $\times$ 8.5 arcmin$^{2}$~and an image scale of
0.5\arcsec\ per pixel. Nine exposures through the H$\alpha$+[N {\sc
ii}] and \oiii\ filters each of 2400 s and nine corresponding
exposures in the continuum, each of 180 s, were taken. During the
observations, the ``seeing'' was varying between 0.8\arcsec and
1.5\arcsec while the full width at half maximum (FWHM) of the star
images was between 1.2\arcsec and 2.1\arcsec. The continuum--subtracted
mosaic of the H$\alpha$+[N {\sc ii}] and \oiii\ images are shown in
Figs. \ref{fig3} and \ref{fig4}, respectively.

\subsection{Spectroscopy}
The 1.3 m Ritchey--Cretien (f/7.7) telescope at Skinakas Observatory
was used to obtain low dispersion long--slit spectra at June 4 and 5
and September 7, 2005. The exposure time was 3900 s, except for G
15.1--1.6 north area, which was 7800 s. The data were taking with a
1300 lines mm $^{-1}$ ~grating and a 2000 $\times$ 800 (13 $\mu$m)
SITe CCD covering the range 4750\AA\ -- 6815\AA. The spectral
resolution is $\sim$8 pixels and $\sim$11 pixels full width at half
maximum (FWHM) for the red and blue wavelengths, respectively. The
airmass of the object was varying between 1.6 and 1.7. The slit has a
width of 7\farcs7 and length of 7\farcm9 and in all cases was oriented
in the south--north direction. The coordinates of the slit centers of
each spectrum are given in Table 1. For the absolute calibration the
spectrophotometric standard stars HR4468, HR5501, HR7596,HR7950,
HR8634 and HR9087 were used. The data reduction was performed by using
the IRAF package.

%
\section{Results}

\subsection{The \hnii, \sii\ and \oiii\ emission line images}
Optical filamentary and diffuse emission is detected for the first time
for this remnant with several thin and curved filaments found all
around the remnant. The most interesting regions lie in the
north--west, west and south--east where complex filamentary structures
exist. In Table \ref{fluxes}, we present typical average fluxes
measured in several locations within the field of G 15.1$-$1.6.  The
detected \sii\ emission appears more diffuse and less filamentary than
in the \hnii\ image, however, its morphology is similar to that in the
\hnii\ therefore, it is not shown here.

Starting from the north, a bright filament
2\arcmin\ long (named A in Fig. \ref{fig1}) is present with its center
approximately at $\alpha \simeq$ 18\h24\m04\s\ and $\delta \simeq$
--16\degr25\arcmin47\arcsec. This filament lies a few arcminutes to
the north--east of the very bright 4\arcmin\ filamentary structure
(named B) which is between $\alpha \simeq$ 18\h23\m49\s, $\delta
\simeq$ --16\degr27\arcmin26\arcsec\ and $\alpha \simeq$ 18\h23\m36\s,
$\delta \simeq$ --16\degr30\arcmin55\arcsec. In particular, there is a
very bright filament 3\arcmin\ long and 1.2\arcmin\ wide with
$\sim$1\arcmin diffuse emission in its south. South--west of this
filament appears a fainter one (C1) at $\alpha \simeq$ 18\h23\m34\s,
$\delta \simeq$ --16\degr35\arcmin08\arcsec\ which is up to 1\arcmin\
long separated by a 2.5\arcmin\ gap with area B. This gap is due to
the existence of a dark region in the area (probably created by dust
-- supported also by the IRAS map) preventing the detection
of optical emission from the SNR. Further to
the south, there is a prominent bright structure which appears also
strong and is designated as filament C2. This structure 
($\sim$3\arcmin\ long,
$\sim$40\arcsec\ wide) lies at $\alpha \simeq$ 18\h23\m30\s, $\delta
\simeq$ --16\degr41\arcmin56\arcsec.  To the south, there is fainter
1\arcmin\ long emission (named D) which has strong \sii\ emission like the
filament in area A (\sii/\ha\ $\sim$0.7). The east structure consists
of two main parts; a very bright one which covers an area of
$\sim$4$\times$4 arcmin$^{2}$, centered at $\alpha \simeq$
18\h24\m17\s, $\delta \simeq$ --16\degr39\arcmin36\arcsec\ (E1) and a
more complex but less bright between $\alpha \simeq$ 18\h24\m38\s,
$\delta \simeq$ --16\degr31\arcmin28\arcsec\ and $\alpha \simeq$
18\h24\m20\s, $\delta \simeq$ --16\degr37\arcmin51\arcsec\
(E2). Diffuse emission is also present close to the filamentary
structures as well as the centre of SNR. All filamentary
structures have the same curvature which supports that they all belong
to G 15.1$-$1.6. It is interesting to note the existence of two very
thin long filaments to the north of E2 extending for
$\sim$10\arcmin\ and join to a 2.5\arcmin single filament. Similar
filaments also apear to the north--east of area A. Both seem to
follow the infrared emission but the low resolution of the latter do
not allow a detailed investigation.

The detected \oiii\ emission (Figs. \ref{fig2}, \ref{fig4}) appears
less filamentary and more diffuse than in the \hnii\ image. In
Table~\ref{fluxes} typical \oiii\ fluxes are listed. Significant
differences between the \hnii\ and \oiii\ images are present for many
of the filaments. In particular, to the north, west and south areas
(A, B, C and D) in contrast to the bright filaments found in \hnii,
the \oiii\ displays a different morphology with much fainter diffuse
emission.  Only within area B, there is a bright \oiii\ filament
centred at $\alpha \simeq$ 18\h23\m43\s, $\delta \simeq$
--16\degr28\arcmin30\arcsec\ which lies exactly at the same position
with the brigth one in \hnii. Further to the west, there is a bright
\oiii\ filament (named F) separated to a very thin one
($\sim$30\arcsec wide) at $\alpha \simeq$ 18\h23\m10\s, $\delta
\simeq$ --16\degr31\arcmin25\arcsec\ and a wider complex structure at
$\alpha \simeq$ 18\h22\m59\s, $\delta \simeq$
--16\degr35\arcmin49\arcsec. This filament does not have \hnii\
counterpart and at the same location only very faint diffuse emission
is found. However, it is not correlated with the 4850 MHz radio map
(Fig. \ref{fig6}) of G 15.1$-1.6$ and probably does not belong to
remnant. On the other hand, the bright \hnii\ emission found to the
east (area E) appears also bright but less filamentary in \oiii.
Finally, a similar situation appears both in \oiii\ and \hnii\ in the
areas where weak and diffuse emission is found.

All images being flux calibrated provide a first indication of the
nature of the observed emission. An examination of the diagnostic
ratio \sii/\ha\ shows that the emission from the brightest parts of
the remnant originates from shock--heated gas since we estimate ratios
\sii/\ha\ of 0.4--0.6, which are in agreement with our spectral
measurements (Sect. 3.2). The north and south areas (A and D) show
\sii/\ha$\sim$0.7. A photoionization mechanism may be producing the 
emission in the south--east region since the ratio \sii/\ha\ is $\sim$0.3.
The possibility
that an \HII\ emission contaminates the remnant's emission to the east
(E) can not be ruled out since for some of the areas close to the
filament as well as to the central region of the remnant, we estimate
\sulfur/\ha\ $\sim$0.3--0.4 which is also in agreement with our
spectra.

Assuming that all the bright filaments belong to the remnant, their
geometry allows us to approximately define its diameter. In
particular, the east border of the remnant is defined by the outer
filament at \a\ $\simeq$ 18\h24\m40.0\s, the west border at \a\
$\simeq$ 18\h23\m25.0\s, the north border at \dd\ $\simeq$
--16\degr24\arcmin\ and the south border at \dd\ $\simeq$
--16\degr46\arcmin. Then a diameter of 30\arcmin $\times$ 24\arcmin
can be derived with its center at \a\ $\simeq$ 18\h24\m00\s, \dd\
$\simeq$ --16\degr35\arcmin20\arcsec. The optically derived
angular size is identical to that quoted in the Green's catalogue
(Green \cite{gre06}).

\subsection{The optical spectra from G 15.1$-$1.6}
The deep low resolution spectra were taken on the relatively bright
optical filaments at two different locations (Table~\ref{table1}). In
Table~\ref{sfluxes} the relative line fluxes taken from the above
locations (designated Area B and E) are quoted. In particular, in Area
B, we extracted two different apertures (BI and BII) along the slit
that are free of field stars and include sufficient line emission to
allow an accurate determination of the observed lines. The background
extraction apertures were selected towards the north and south ends of
each slit depending on the filament's position within the slit. The
measured line fluxes indicate emission from shock heated gas, since
\sii/\ha\ $\simeq$0.5. Furthermore, the \nii/\ha\ ratio, which takes
values between 0.63 and 1.07 (see Table~\ref{sfluxes}, falls well
inside the range expected for a SNR (Fesen et al. \cite{fes85}). The
signal to noise ratios do not include calibration errors, which are
less than 10 percent. Typical spectra from the north (BI) and
south (E1) areas are shown in Fig. 5.  
\par 
The absolute \ha\ flux covers a range of values, from 2 to 7
$\times$ \fluxa. The \siirat\ ratio which was calculated between 1.3
and 1.4, indicates electron densities between 40 to 150 cm$^{-3}$~
(Osterbrock \& Ferland \cite{ost06}). However, taking into account the
statistical errors on the sulfur lines, we calculate that electron
densities up to 250 cm$^{-3}$ are allowed (Shaw \& Dufour
\cite{sha95}). Measurements for the \oiii/\hbeta\ ratio result in
values less than 6. Theoretical models of Cox \& Raymond
(\cite{cox85}) and Hartigan \et\ (\cite{har87}) suggest that for
shocks with complete recombination zones this value is $\sim$6, while
this limit is exceeded in case of shock with incomplete recombination
zones (Raymond \et\ \cite{ray88}). Our measured values suggest that
shocks with complete recombination zones are present. Therefore,
according to our measurements and the above theoretical models the
estimated shock velocities are $\sim$100 \vel.

\subsection{Observations at other wavelengths}

The optical emission matches very well the radio emission of G
15.1$-$1.6 at 4850 MHz, suggesting their correlation
(Fig.~\ref{fig6}). The observed filaments are located close to the
outer edge of the radio contours but the low resolution of the radio
images does not allow us to determine the relative position of the
filament with respect to the shock front. In order to explore how the
optical emission correlates with the infrared emission, IRAS images
at 60 $\mu$m of the same area were examined. Fig. 6 shows a deep
greyscale representation of the optical emission (\hnii) with
overlapping contours of the infrared emission (60 $\mu$m). Although
the low-resolution IRAS map does not permit a detailed comparison
with the optical image there is a clear enhancement of infrared
emission in the area where the optical emission of the SNR is
detected.  The infrared emission closely follows the morphology of the
SNR but it also fills the central area which is empty of optical
emission. We have also examined the ROSAT All--sky survey data but 
no significant X--ray emission was detected.

%

\section{Discussion}
The \snr\ G 15.1--1.6 shows up as an almost complete shell in the
radio band without any X--ray emission detected so far. The absence of
soft X--ray emission may indicate a low shock temperature and/or a low
density of the local interstellar medium. The \hnii\ image best
describe the newly detected structures. \sii\ and \oiii\ emission is
also detected and generally appears less filamentary and more diffuse
than in the \hnii\ image with their position and shape to be in
agreement with that of the \hnii. The \oiii\ emission seems not to be
bounded by \hnii\ emission. An explanation is given by Blair et
al. (\cite{bla05}), indicating that the shock emission from the
nascent radiative region is sufficient to fully ionize the local
preshock gas. The presence of [O {\sc i}] 6300 \AA\ line emission
is also consistent with the emission being shock material. Both the
calibrated images and the long--slit spectra suggest that the detected
emission results from shock heated gas since the \sii/\ha\ ratio
exceeds the empirical SNR criterion value of 0.4--0.5, while the
measured \nii/\ha\ ratio also confirms this result.  Note that the
possible \HII\ region found in the low ionization images shows a
\sii/\ha\ ratio of $\sim$0.33. The eastern filament lies very close to
this \HII\ region. The morphological differences between the low and
medium ionization lines provide evidence for significant
inhomogeneities and density variations in the ambient medium. Hester
et al. (1987) suggested that the presence of such inhomogeneities and
density variations would mainly affect the recombination zone where
the low ionization lines are produced and it could also explain the
\oiii/\ha\ ratio variations seen in the long--slit spectra.
\par
An interstellar extinction c between (see Table~\ref{sfluxes}) 1.17
($\pm$ 0.05) and 1.68 ($\pm$ 0.05) or an A$_{\rm V}$~between 2.51
($\pm$ 0.11) and 3.60 ($\pm$ 0.10) were measured, respectively. We
have also determined the electron density measuring the density
sensitive line ratio of \siirat. The densities we measure are below
250 \dens. Assuming that the temperature is close to 10$^{4}$ K, it is
possible to estimate basic SNR parameters. The remnant under
investigation is one of the least studied remnants and thus, the
current stage of its evolution is unknown. Assuming that the remnant
is still in the adiabatic phase of its evolution the preshock cloud
density n$_{\rm c}$ can be measured by using the relationship (Dopita
\cite{dop79})

\begin{equation}
{\rm n_{[SII]} \simeq\ 45\ n_c V_{\rm s}^2}~{\rm cm^{-3}},
\end{equation}

where ${\rm n_{[SII]}}$ is the electron density derived from the
sulfur line ratio and V$_{\rm s}$ is the shock velocity into the
clouds in units of 100 \vel. Furthermore, the blast wave energy can be
expressed in terms of the cloud parameters by using the equation given
by McKee \& Cowie (\cite{mck75})

\begin{equation}
{\rm E_{51}} = 2 \times 10^{-5} \beta^{-1} 
{\rm n_c}\ V_{\rm s}^2 \ 
{\rm r_{s}}^3 \ \ {\rm erg}.
\end{equation}

The factor $\beta$ is approximately equal to 1 at the blast wave
shock, ${\rm E_{51}}$ is the explosion energy in units of 10$^{51}$
erg and {\rm r$_{\rm s}$} the radius of the remnant in pc. By using
the upper limit on the electron density of 250 \dens, which was
derived from our spectra, we obtain from Eq. (1) that ${\rm n_c}
V_{\rm s}^2 < 5.6$. Then Eq. (2) becomes ${\rm E_{51}} < 9 \times
10^{-3}~{\rm D_{1 kpc}^3}$, where ${\rm D_{1 kpc}}$~the distance to
the remnant in units of 1 kpc.
\par
Estimated values of N$_{\rm H} \sim 6.6 \times 10^{21}$~cm$^{-2}$ and
N$_{\rm H} \sim 8.3 \times 10^{21}$~cm$^{-2}$ are given by Dickey \&
Lockman (\cite{dic90}) and Kalberla et al. (\cite{kal05})
respectively, for the column density in the direction of G
15.1$-$1.6. Using the relation of Ryter et al. (\cite{ryt75}), we obtain
an N$_{{\rm H}}$~of $5.5 \times 10^{21}~{\rm cm}^{-2}$~and $7.9 \times
10^{21}~{\rm cm}^{-2}$~for the minimum and maximum c values calculated
from our spectra, respectively. Both values are consistent with the
estimated galactic N$_{\rm H}$~considering the uncertainties involved.
Since, there are no other measurements of the interstellar density
n$_{0}$, values of 0.1 and 1.0 will be examined. Following the result
of Eq. (2) and assuming the typical value of 1 for the supernova
explosion energy (E$_{51}$), we find that the remnant may lie at
distance greater than 2.2 kpc. Then, the lower interstellar density
of $\sim$0.1 cm$^{-3}$~suggests that the column density is greater
than $1.4 \times 10^{21}~{\rm cm}^{-2}$, while for n$_{0} \approx
1~{\rm cm}^{-3}$~it becomes greater than $1.4 \times 10^{22}~{\rm
cm}^{-2}$. Combining the previous results and assuming that the column
density is found in the range of $5 - 8 \times 10^{21}~{\rm cm}^{-2}$,
then the lower interstellar density seems to be more
probable. However, since neither the distance nor the interstellar
medium density are accurately known, we cannot confidently determine
the current stage of evolution of G 15.1$-$1.6.


\section{Conclusions}
The faint supernova remnant G 15.1$-$1.6 was observed for the first
time in major optical emission lines. The images show filamentary and
diffuse emission structures. The bright filaments are very well
correlated with the remnant's radio emission at 4850 MHz suggesting
their association, while correlation evidence also shown with the IRAS
60$\mu$m map. The flux calibrated images and the long--slit spectra
indicate that the emission arises from shock heated gas. Finally, an
upper limit for the electron density of 250 \dens and a lower limit
for the distance of 2.2 kpc are calculated.


\begin{acknowledgements}

JA acknowledges funding by the European Union and the Greek Ministry
of Development in the framework of the programme `Promotion of
Excellence in Research Institutes (2nd Part)'. Skinakas Observatory is
a collaborative project of the University of Crete, the Foundation for
Research and Technology-Hellas and the Max-Planck-Institut f\"ur
Extraterrestrische Physik.
\end{acknowledgements}

\newpage
%

%

\begin{table*}  
\caption[]{Imaging and Spectral log}  
\label{table1}
\begin{flushleft} 
\begin{tabular}{lcccccc}  
\noalign{\smallskip}  
\hline  
\multicolumn{6}{c}{IMAGING} \\  
\hline
Filter & $\lambda_{\rm c}$ & $\Delta \lambda$ & Total exp. time &
N$^{\rm o}$ of diff. fields & Telescope\\ & ($\AA$) & ($\AA$) & (sec)
& & \\
\hline
\HNII & 6570 & 75 & 7200 (3)$^{\rm a}$ & 1 & 0.3-m \\
\OIII & 5010 & 28 & 7200 (3) & 1 & 0.3-m \\
\SII & 6720 & 18 & 4800 (2) & 1 & 0.3-m \\
Cont blue & 5470 & 230 & 180 & 1 & 0.3-m \\
Cont red & 6096 & 134 & 180 & 1 & 0.3-m \\
\OIII & 5010 & 28 & 2400 & 9 & 1.3-m \\
Cont blue & 5470 & 230 & 180 & 9 & 1.3-m \\
\HNII & 6570 & 75 & 2400 & 9 & 1.3-m \\
Cont red & 6096 & 134 & 180 & 9 & 1.3-m \\
\hline
\multicolumn{7}{c}{SPECTROSCOPY} \\  
\hline  
Area & \multicolumn{2}{c}{Slit centres} & Total exp. time &
Offset$^{\rm b}$ & Aperture length$^{\rm c}$ & Telescope \\  
 & $\alpha$ & $\delta$ & (sec) & (arcsec) & (arcsec)& \\  
\hline  
North (BI) & 18\h23\m44.0\s & -16\degr29\arcmin05\arcsec & 7800 (2)
& 10.6 N & 38.9 & 1.3-m \\
North (BII) & 18\h23\m44.0\s & -16\degr29\arcmin05\arcsec & 7800
 (2) & 96.8 N & 13.0 & 1.3-m\\
South--East (E1) & 18\h24\m14.6\s & -16\degr36\arcmin38\arcsec & 3900 &
 123.3 S & 15.3 & 1.3-m\\
\hline  
\end{tabular}
\end{flushleft}
\begin{flushleft}
${\rm ^a}$ Numbers in parentheses represent the number of individual frames.\\
${\rm ^b}$ Spatial offset from the slit centre in arcsec: N($=$North),
S($=$South).\\ 
${\rm ^c}$ Aperture lengths for each area in arcsec.\\
\end{flushleft} 
\end{table*}

\begin{table}
\caption[]{Typically measured fluxes over the brightest filaments.}
\label{fluxes}
\begin{flushleft}
\begin{tabular}{lllllll}
\hline
\noalign{\smallskip}
 & A & B & C & D & E & F$^{\rm a}$ \\
\hline
\hnii\  & 86 & 156 & 111 & 80 & 152 & 24 \\
\hline
\sii\   & 30 & 38 & 30 & 28 & 37 & $<$5$^{\rm b}$\\
\hline
\oiii\  & 12 & 5 & 20 & 9 & 16 & 21 \\
\hline
\end{tabular}
\end{flushleft}
${\rm }$ Fluxes in units of \fluxb \\
${\rm}$ Median values over a 40\arcsec $\times$ 40\arcsec\ box. \\
${\rm ^a}$ The west \oiii\ bright filament outside the snr's borders.\\ 
${\rm ^b}$ 3$\sigma$~upper limit.\\
 \end{table}

\begin{table*}
\caption[]{Relative line fluxes.}
\label{sfluxes}
\begin{flushleft}
\begin{tabular}{llllllllll}
\hline \noalign{\smallskip} & \multicolumn{3}{c}{Area BI} &
\multicolumn{3}{c}{Area BII} & \multicolumn{3}{ c}{Area E} \\ 
Line (\AA) & F$^{\rm a}$ & I$^{\rm b}$ & S/N$^{\rm c}$ & F & I & S/N & F &
I & S/N \\ 
\hline \hbeta\ 4861 & 9 & 35 & (27) & 11 & 35 & (11) & 13 & 35 & (25) \\ 
\oxygen\ 4959 & 6 & 20 & (20) & 11 & 31 & (12) & 7 & 18 & (17) \\
\oxygen\ 5007 & 7 & 29 & (33) & 14 & 39 & (18) & 18 & 43 & (39) \\ 
\oi\ 6300 & 4 & 5 & (35) & 6 & 7 & (14) & 6 & 7 & (33) \\ 
\nitrogen\ 6548 & 26 & 26 & (160) & 20 & 20 & (44) & 16 & 15 & (74) \\ 
\ha\ 6563 & 100 & 100 & (441) & 100 & 100 & (162) & 100 & 100 & (276) \\ 
\nitrogen\ 6583 & 81 & 79 & (373) & 63 & 62 & (109) & 47 & 47 & (163) \\ 
\sulfur\ 6716 & 30 & 27 & (161) & 28 & 26 & (52) & 19 & 18 & (81) \\ 
\sulfur\ 6731 & 24 & 21 & (129) & 22 & 20 & (41) & 14 & 13 & (62) \\
\hline Absolute \ha\ flux$^{\rm d}$ & \multicolumn{3}{c}{3} &
\multicolumn{3}{c}{2} & \multicolumn{3}{c}{7} \\
\sulfur/\ha\ & \multicolumn{3}{c}{0.50 $\pm$ 0.03} &
\multicolumn{3}{c}{0.48 $\pm$ 0.07} & \multicolumn{3}{c}{0.33 $\pm$
0.02} \\ 
F(6716)/F(6731) & \multicolumn{3}{c}{1.29 $\pm$ 0.10} &
\multicolumn{3}{c}{1.32 $\pm$ 0.25} & \multicolumn{3}{c}{1.38 $\pm$
0.12} \\
\nii/\ha\ & \multicolumn{3}{c}{1.07 $\pm$ 0.02} &
\multicolumn{3}{c}{0.83 $\pm$ 0.04} & \multicolumn{3}{c}{0.63 $\pm$
0.01} \\ 
 \oxygen/\hbeta\ & \multicolumn{3}{c}{1.40 $\pm$ 0.09} &
\multicolumn{3}{c}{2.00 $\pm$ 0.30} & \multicolumn{3}{c}{1.73 $\pm$
0.12} \\ 
c(\hbeta) & \multicolumn{3}{c}{1.68 $\pm$ 0.05} &
\multicolumn{3}{c}{1.38 $\pm$ 0.11} & \multicolumn{3}{c}{1.17 $\pm$
0.05} \\ 
E$_{\rm B-V}$ & \multicolumn{3}{c}{1.16$\pm$0.03} &
\multicolumn{3}{c}{0.95$\pm$ 0.08} & \multicolumn{3}{c}{0.81$\pm$0.03}
\\ 
\hline
\end{tabular}
\end{flushleft}  
\begin{flushleft}
${\rm ^a}$ Observed fluxes normalised to F(H$\alpha$)=100 and uncorrected  
for interstellar extinction.  

${\rm ^b}$ Intrinsic fluxes normalised to F(H$\alpha$)=100 and corrected  
for interstellar extinction. 

${\rm ^c}$ Numbers in parentheses represent the signal to noise ratio of the
quoted fluxes.\\ 

$^{\rm d}$ In units of \fluxa.  

Listed fluxes are a signal to noise weighted average of two fluxes for
areas BI and BII.\\

The emission line ratios \sulfur/\ha, F(6716)/F(6731) and
\oxygen/\hbeta\ are calculated using the corrected for interstellar
extinction values.\\

The errors of the emission line ratios, c(\hbeta) and E$_{\rm B-V}$
are calculated through standard error propagation.\\
\end{flushleft}   

\end{table*}

\newpage

\begin{figure*}
\centering
\includegraphics[bb=0 0 519 651]{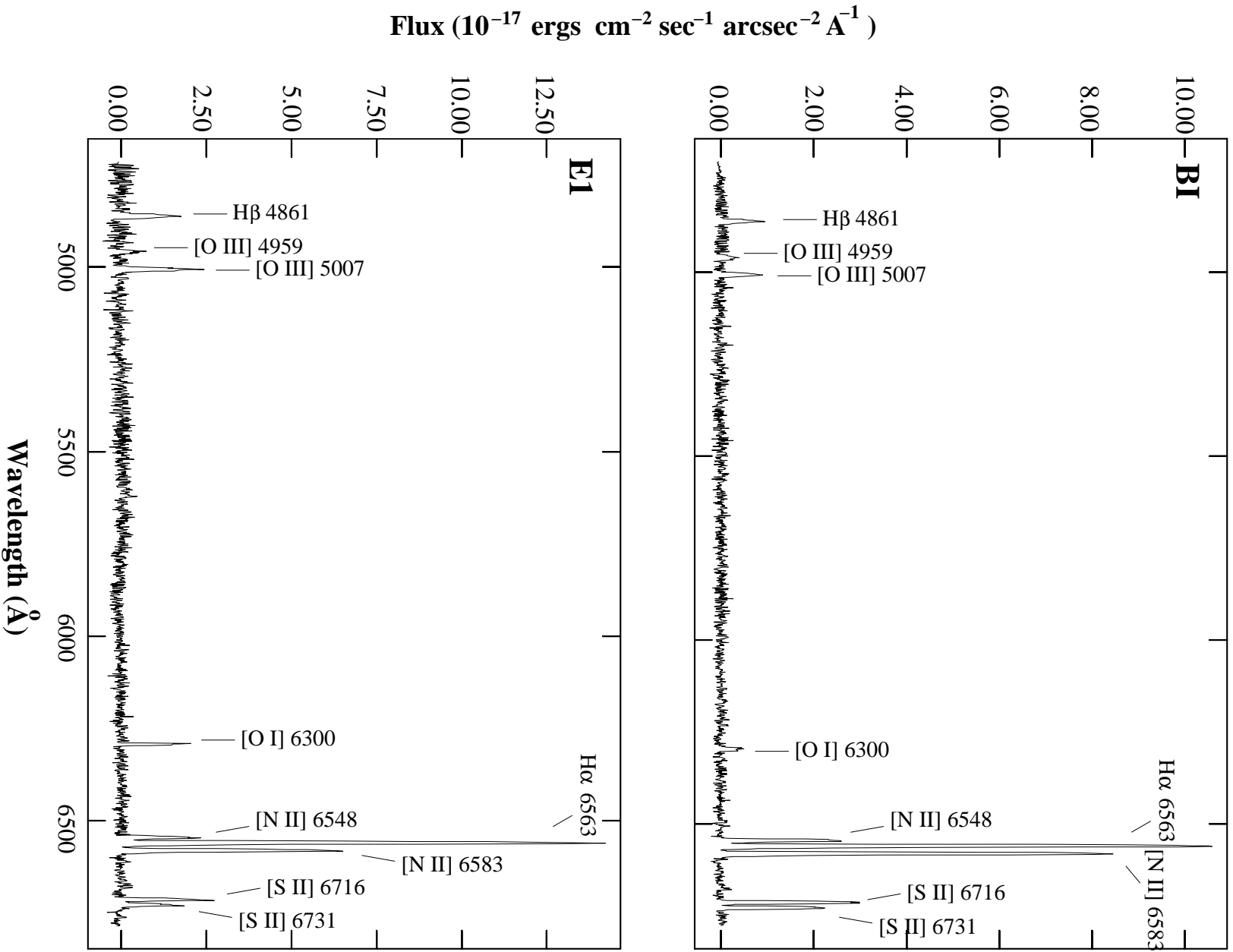}
\caption{Typical long--slit spectra.}
\label{fig5}
\end{figure*}

\begin{figure*}
\centering
\includegraphics[bb=150 0 719 551]{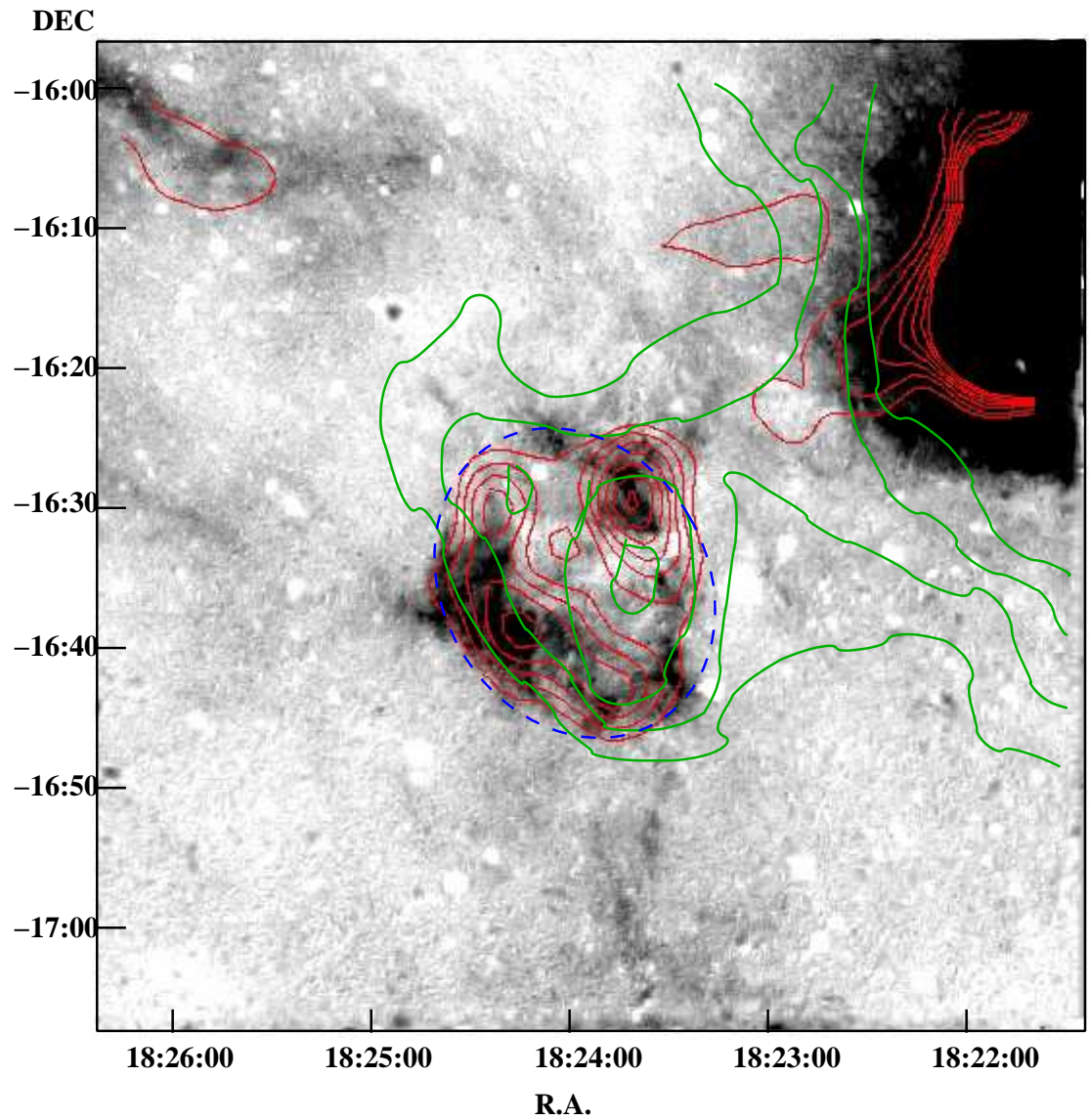}
\caption{The correlation between the H$\alpha +$[N {\sc ii}] emission,
the radio emission at 4850 MHz (solid red line), the infrared emission
at IRAS 60 $\mu$m (solid green line) is shown in this figure. The 4850
MHz radio contours scale linearly from 3.54$\times 10^{-2}$ Jy/beam to
0.3 Jy/beam while the 60 $\mu$m contours are at the level of 170, 200,
230 and 260 MJy/sr. The ellipse (dash blue line) approximately defines
the SNR's boundary. }
\label{fig6}
\end{figure*}

\end{document}